\documentclass[fleqn,10pt]{wlscirep}
\usepackage{amsmath,bm}
\usepackage{graphicx}
\def\be{\begin{eqnarray}}
\def\ee{\end{eqnarray}}
\def\r{{\bf r}}

\def\E{{\bf E}}

\def\p{{\bf p}}

\def\k{{\bf k}}

\def\G{{\bf G}}

\def\ep{{\bm{\epsilon}}}

\title{Magneto-Optical Activity in High Index Dielectric Nanoantennas}

\author[1,2]{N. de Sousa}
\author[3]{L. S. Froufe-P\'erez}
\author[2,4,*]{J. J. S\'aenz}
\author[5,+]{A. Garc\'ia-Mart\'in}

\affil[1]{Departamento de F\'isica de la Materia Condensada,  Condensed Matter Physics Center (IFIMAC) and Instituto ``Nicol\'{a}s Cabrera'', Universidad Aut\'{o}noma de Madrid, 28049, Madrid, Spain.}
\affil[2]{Donostia International Physics Center (DIPC), P. Manuel de Lardizabal
4, Donostia-San Sebasti\'{a}n 20018, Spain.}
\affil[3]{Department of Physics, University of Fribourg, Chemin du Mus\'ee 3, CH-1700 Fribourg, Switzerland.}
\affil[4]{IKERBASQUE, Basque Foundation for Science, 48013 Bilbao, Spain.}
\affil[5]{IMM - Instituto de Microelectr\'{o}nica de Madrid (CNM-CSIC), Isaac Newton 8, PTM, Tres Cantos, E-28760 Madrid, Spain.}

\affil[*]{juanjo.saenz@dipc.org}

\affil[+]{a.garcia.martin@csic.es}


\begin{abstract}
The magneto-optical activity, namely the polarization conversion capabilities of high-index, non-absorbing, core-shell dielectric nanospheres is theoretically analyzed. We show that, in analogy with their plasmonic counterparts, the polarization conversion in resonant dielectric particles is linked to the amount of electromagnetic field probing the magneto-optical material in the system. However, in strong contrast with plasmon nanoparticles, due to the peculiar distribution of the internal fields in resonant dielectric spheres, the magneto-optical response is fully governed by the magnetic (dipolar and quadrupolar) resonances with little effect of the electric ones. 
\end{abstract}
\begin{document}

\flushbottom
\maketitle

\thispagestyle{empty}


\section*{Introduction}

The ability to externally control the propagation of light in the visible and near-infrared domain by means nanostructured materials has been a matter of intense research in the last decade. This interest is explained by the promising potential applications in different areas of technology, like telecommunications \cite{temnov2010active, belotelov2011enhanced} or sensing \cite{sepulveda2006highly,caballero2016hybrid}. A way to modify the scattered light, such as intensity, directionality, phase and polarization is by using small metallic particles compared with the wavelength. The interaction of light with these particles, usually referred as nanoantennas, can be moulded by changing their characteristics such as size, material or shape \cite{link1999spectral, kelly2003optical, aizpurua2003optical, lee2006gold, bryant2008mapping, rodriguez2012gold}. 
This is driven by the possibility to excite localized surface plasmons and the subsequent strong near field interactions allow the fabrication of systems with high directionality \cite{taminiau2008enhanced, novotny2011antennas} or obtain configurations where the electromagnetic field is confined in small volumes \cite{gonzalez2008plasmonic}.\\
In the quest to  exert certain degree of control of the plasmon properties using external parameters, the so-called active plasmonics, some developments have been made using different  "controlling agents". Electric fields \cite{dicken2008electrooptic}, temperature \cite{nikolajsen2004surface} or electromagnetic waves \cite{pacifici2007all} have been used as such external agents. An interesting alternative is the use of an external magnetic field \cite{martin2010enhancement,armelles2013magnetoplasmonics}. In this case, the reverse effect is also very interesting, namely, to use the plasmon resonance to enhance the magneto-optical response \cite{maccaferri2015ultrasensitive, lodewijks2014magnetoplasmonic, maccaferri2013polarizability}. An important aspect in this case is that the internal architecture of the plasmonic elements can largely modify the way that the enhancement is realized\cite{meneses2011probing,armelles2013mimicking, de2014interaction}, since the actual distribution of the electromagnetic field in the material plays a crucial role\cite{meneses2011probing,rollinger2016localizaton}. \\
In the last years the concept of optical magnetic resonances in the visible domain has been put forward for its evident interest in terms of scattering efficiency,\cite{alu2009quest} and magneto-optical response\cite{Armelles_2015_Nanoletters}, this last based on the Babinet principle for plasmonic entities\cite{Mole_babinet1,Mole_babinet2}.

On the other hand, dielectric materials present themselves as a particularly interesting alternative to resonant dipolar-like scattering elements. High refractive dielectric nanoparticles were shown to present both strong electric and magnetic dipolar resonances  \cite{Evlyukhin2010,garcia2011strong}  exhibiting weak dissipation in the visible and practically lossless in telecomm and near-infrared frequencies. 
Linked to these properties, an increasing interest in the use of high index dielectric nanoparticles as optical antennas has emerged \cite{gomez2011electric,geffrin2012magnetic,krasnok2012all,schmidt2012dielectric,Rolly2012,marinchio2014njp,tribelsky2015small,paniagua2016natcom}.   

In this paper we address the magneto-optical effect in the context of these high index dielectric nano-antennas. To illustrate the effect we will consider a practical case where the antenna is a silicon nanosphere with non-negligible off-diagonal elements in the dielectric tensor. We show that, as expected, the magneto-optical effect is controlled by the internal resonances of the nanosphere, but, contrary to the case of metals where electric dipoles dominate, the magnetic resonances are the ones that dominate the spectral dependence of the magneto-optical response, having the electric dipolar resonance a small, even tiny, influence. Additionally, we establish a clear correlation of the spectral magneto-optical response with that of the spatial field profile within the nanosphere that is, in turn, linked to the nature of each resonance.

\section*{Results and discussion}


\begin{figure}[ht]
\centering
\includegraphics[width=10cm]{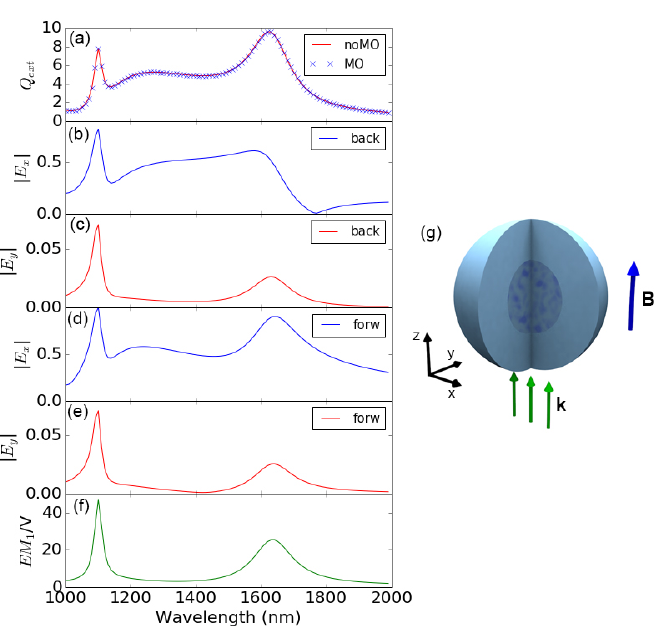}
\caption{(a) Extinction efficiency of the Silicon nanoparticle with $230$nm of radius, without external magnetic field (continuous line) and with it (symbols); (b) and (c) spectral dependence of the $x$ and $y$ component of the electromagnetic field in the backscattering direction; (d) and (e) spectral dependence of the $x$ and $y$ component of the electromagnetic field in the forward direction; all fields in (b)-(e) have been normalized to the maximum value, occurring for  $E_x^{\text{forw}}$. (f) Integral of the electromagnetic field intensity inside the sphere, normalized to the sphere volume. (g) sketch of the spherical nanoantenna with a MO core depicting  the incoming field and the orientation of the external static magnetic field. 
}
\label{fig:1}
\end{figure}

Our model system will be a high index, non-absorbing, dielectric nanoantenna consisting of a spherical particle with radius 230nm made of Silicon (n=3.5). This nanoantenna is further illuminated by a plane wave (with intensity $E_0$) impinging along the $z-$axis and with its polarization aligned along the $x-$direction. The Silicon particle is assumed to be a core-shell where the core is uniformly doped within a  
MO material [see sketch in Fig. \ref{fig:1}]. In the presence of a static magnetic field along $z$ (i.e. in parallel of the incident electromagnetic plane wave)  the dielectric permittivity  of the, otherwise isotropic, material becomes a  non-diagonal tensor of the form 
\begin{equation}
\boldsymbol{\epsilon} = \left(\begin{array}{ccc}
\epsilon_{d} & \epsilon_{xy} (\r)& 0\\
-\epsilon_{xy}(\r) & \epsilon_{d} & 0\\
0 & 0 & \epsilon_{d}
\end{array}\right),\label{eq:dielectric_tensor}
\end{equation}
where $\epsilon_{d}=n^2=12.25$ and $\epsilon_{xy}(\r) = i Q \epsilon_{d} m_z (\r)$, which is proportional to the magnetization along $z$, $m_z$,  accounts for the nonuniform distribution of magnetic impurities in the otherwise homogeneous Si sphere. For simplicity we will assume a lossless response with $\epsilon_{xy}(\r) \sim  0.1 \ i $, a value that is easily achievable using dielectric MO active materials such as garnets [The permittivity tensor is then Hermitian, i.e.  $\boldsymbol{\epsilon} = \boldsymbol{\epsilon}^\dagger$]. 

In order to obtain its electromagnetic response of the nanoantenna to the incident field, we will use an extended  discrete dipole approximation (DDA), where the particle is divided in $\mathcal{N}=4224$ elements with identical volume. 
The explicit expressions that allow calculating the electric field at any point in space, as well as the extinction, absorption, and scattering cross sections can be found in the Methods Section [Eqs. \eqref{pnDDA}-\eqref{abs}].

\begin{figure}[ht]
\centering
\includegraphics[width=15cm]{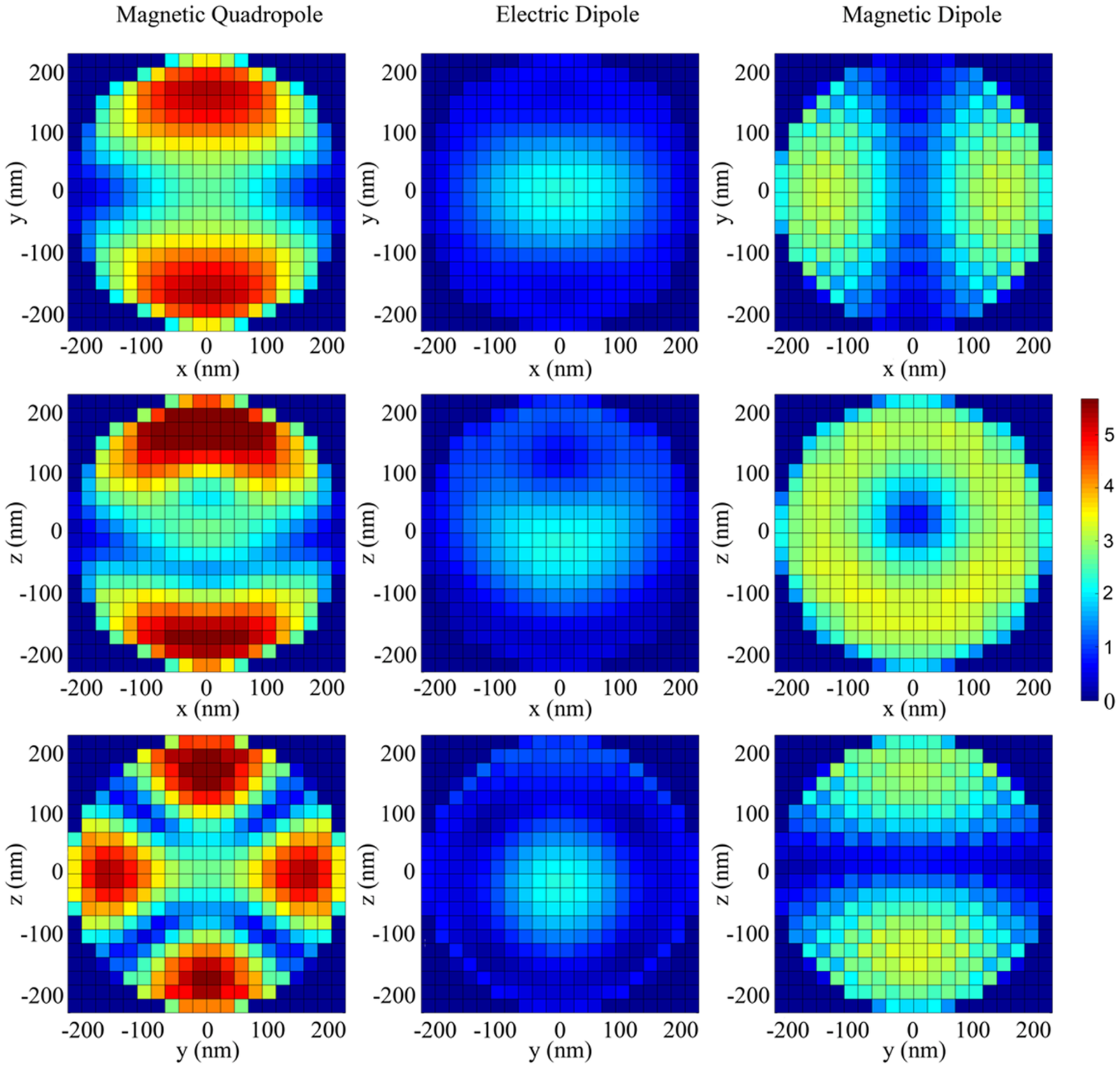}
\caption{Spatial profiles of the EM field norm ($|\mathbf{E}({\mathbf{r}})/E_0|$) inside the sphere in the three principal planes $(x,y, z=0)$, $(x,y=0,z)$ and $(x=0,y,z)$ for frequencies corresponding to the three resonant modes, magnetic quadrupole ($\lambda\approx 1100\text{nm}$), electric dipole ($\lambda\approx 1250\text{nm}$), magnetic dipole ($\lambda\approx 1625\text{nm}$).}
\label{fig:2}
\end{figure}

In Fig. \ref{fig:1}(a) we show the Extinction Efficiency (extinction cross section normalized to the geometrical cross section) of our model system as a function of the wavelength of the impinging radiation. It displays the typical peaked structure corresponding to the most fundamental excitations: magnetic dipole ($\lambda\approx 1625\textnormal{nm}$),
electric dipole ($\lambda\approx 1250\textnormal{nm}$) and magnetic quadrupole ($\lambda\approx 1100\textnormal{nm}$) \cite{garcia2011strong}. In the same figure we present the scattering cross section when a static magnetic field is applied in a way such that the permittivity tensor is MO active and homogeneous affecting equally the whole sphere (i.e. all discretization elements present a dielectric tensor as in \ref{eq:dielectric_tensor}). As it can be seen, and as it is commonly 
the case in magneto-optical effects in the visible and near-IR part of the electromagnetic spectrum, the optical properties, in this case the cross sections, do not experiment a noticeable modification.
However, the presence of the non-diagonal (magneto-optical) elements in the dielectric tensor implies that the polarizability of the sphere is also non-diagonal inducing a  polarization conversion\cite{armelles2013mimicking, de2014interaction}. From the original incoming  x-polarized wave $E_x$, the back-reflected or transmitted wave will have a small, MO-induced y-component $E_y$. This effect is commonly known as the Polar Kerr effect, for reflected waves,  or Polar Faraday effect for transmitted ones. 
Thus in Figs. \ref{fig:1}(b), \ref{fig:1}(c) we present the spectral dependence of the $x$ and $y$ components of the electromagnetic (EM) field in the backscattered direction. $E_x$ presents the distinctive marks of the fundamental excitations (magnetic quadrupole and electric and magnetic dipoles) together with a zero-field point at $\lambda\approx 1775\textnormal{nm}$. This point corresponds to the first Kerker condition, where both electric and magnetic dipolar resonances scatter coherently, leading to a zero-backward field intensity in the radiated power \cite{gomez2011electric,geffrin2012magnetic}. $E_y$, however, does not follow that principle and displays only two well defined peaks, spectrally located at the position of the magnetic dipole and magnetic quadrupole, with a very weak shoulder the position of the electric dipole. This fact is rather striking, since the excitation of electric dipoles is the basis of the extensively addressed enhancement of the MO activity in metallic magnetoplasmonic systems. 
Similarly, in Figs. \ref{fig:1}(d) and \ref{fig:1}(e) the spectral dependence of the $x$ and $y$ components of the EM field in the forward direction is shown. As it can be seen, $E_x$ contains basically the same information as the cross section, with well defined peaks at the fundamental excitations, and preserving the same relative intensities [Notice that from the Optical Theorem, the imaginary part of the amplitude of the forward wave is proportional to the extinction cross section]. On the other hand, $E_y$ is very similar to the that in the backward direction. 
In the context of magneto-optical activity in resonant systems, it has been pointed out that the polarization conversion can be linked to the amount of electromagnetic field probing the magneto-optical material in the system \cite{meneses2011probing, rollinger2016localizaton}. Thus, in Fig. \ref{fig:1}(f), we present the integral of the EM field Intensity ($EM_I=\int |\mathbf{E}({\mathbf{r}})/E_0|^2 d^3r$) inside the sphere, normalized to the sphere volume. As it can readily be seen, the integration reveal only two clear peaks in the spectrum, in almost perfect resemblance of the converted component of the backscattered and/or forward far fields.

To verify this assertion, we present in Fig. \ref{fig:2} the profiles of the EM field norm  ($|\mathbf{E}({\mathbf{r}})/E_0|$) inside the sphere in the three principal symmetry planes $XY=(x,y,z=0)$, $XZ=(x,y=0,z)$ and $YZ=(x=0,y,z)$ for the frequencies corresponding to the three resonant modes. As it can be seen the weakest contribution comes from the region where electric dipole is excited, whereas the strongest is from that where the excited mode is the magnetic quadrupole. This fact accurately coincides with the spectral distribution of the polarization conversion presented in Figs. \ref{fig:1}(c) and (e), and the integrated intensity within the sphere in Fig. \ref{fig:1}(f).  Moreover, the spatial distribution of the intensities points to a higher localization towards the center of the sphere for the case of the electric dipole, being the magnetic dipole basically absent at the sphere center. The case of the quadrupole is more complex, since being weaker at the center, the field  intensity is still larger than that of the electric dipole.

Let us now consider core@shell structures for the MO activity, i.e. the MO activity is only located within a central region of the sphere, while keeping the rest non MO active. For small radii of the MO core, the contribution of the electric dipole would be high, competing with that of the magnetic quadrupole. As the core radius increases, this contribution loses relative weight with respect to the magnetic resonances, which will dominate for bigger radii. For the magnetic dipole the situation should be close to complementary to that of the electric dipole, when the core is small there should be a very weak contribution that increases as the radius of the core grows. We also expect that the strongest contribution will always be from the  from the quadrupole irrespective of the radius of the core, but with varying relative intensities.

\begin{figure}[ht]
\centering
\includegraphics[width=15cm]{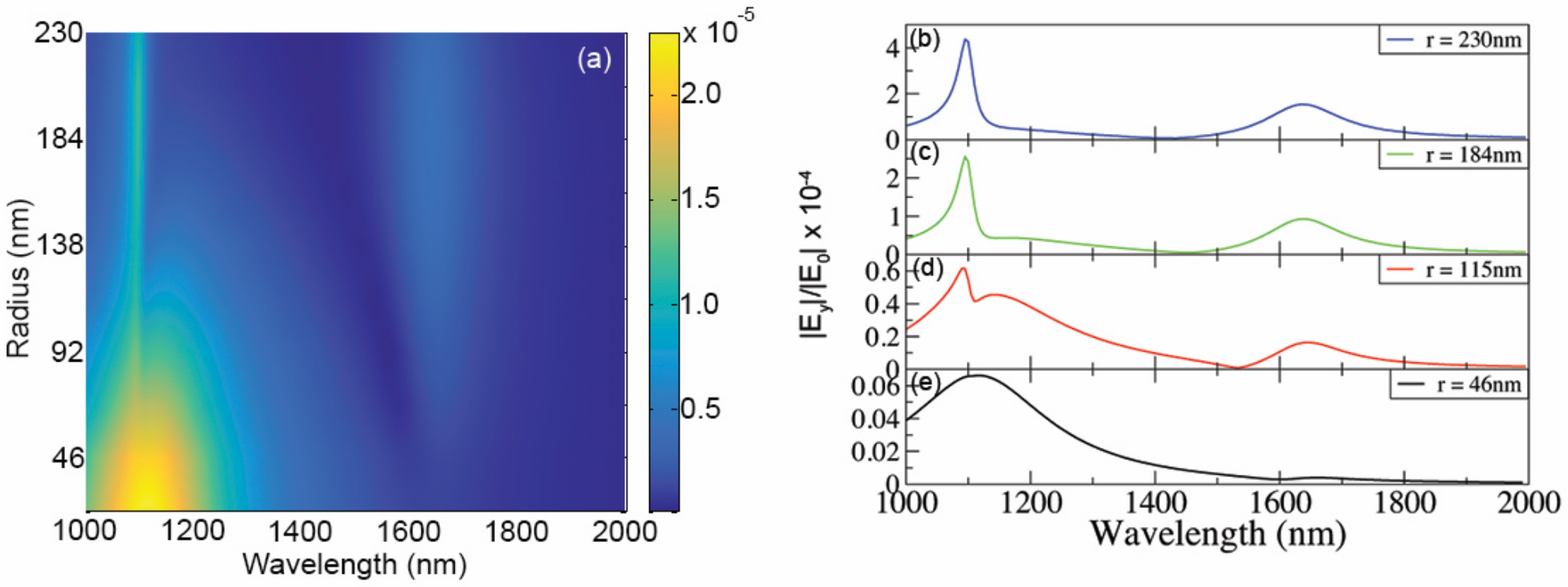}
\caption{(a) Polarization conversion ($|E_y/E_0|$), normalized to the amount of MO material (volume of the core), as a function of both the core radius and the wavelength. Polarization conversion for different core radii: 230nm (b), 184nm (c), 115nm (d) and 46nm (e) showing the evolution of the contribution of the main resonances. }
\label{fig:3}
\end{figure}

This fact is nicely displayed in Fig. \ref{fig:3}(a) where we present the polarization conversion ($|E_y/E_0|$), normalized to the amount of MO material for a better view, as a function of the core radius and of the wavelength. 
As the field profiles of the resonances indicate, for small core radii the largest contribution is spectrally localized at the region where the magnetic quadrupole and electric dipole are excited, being basically non-existent in the region of the magnetic dipole. As the core radius increases the relative contribution of the electric dipole decreases, whereas that of the magnetic resonances increases, being the magnetic quadrupole dominant irrespective of the value of the core radius. Figs. \ref{fig:3}(b)-(e) present the same information for selected radii, displaying in this case the bare, not normalized, intensities. The overall polarization increases as the amount of MO material does, whereas the relative weight is that inferred from the field profiles inside the nanoantenna.

\begin{figure}[ht]
\centering
\includegraphics[width=10cm]{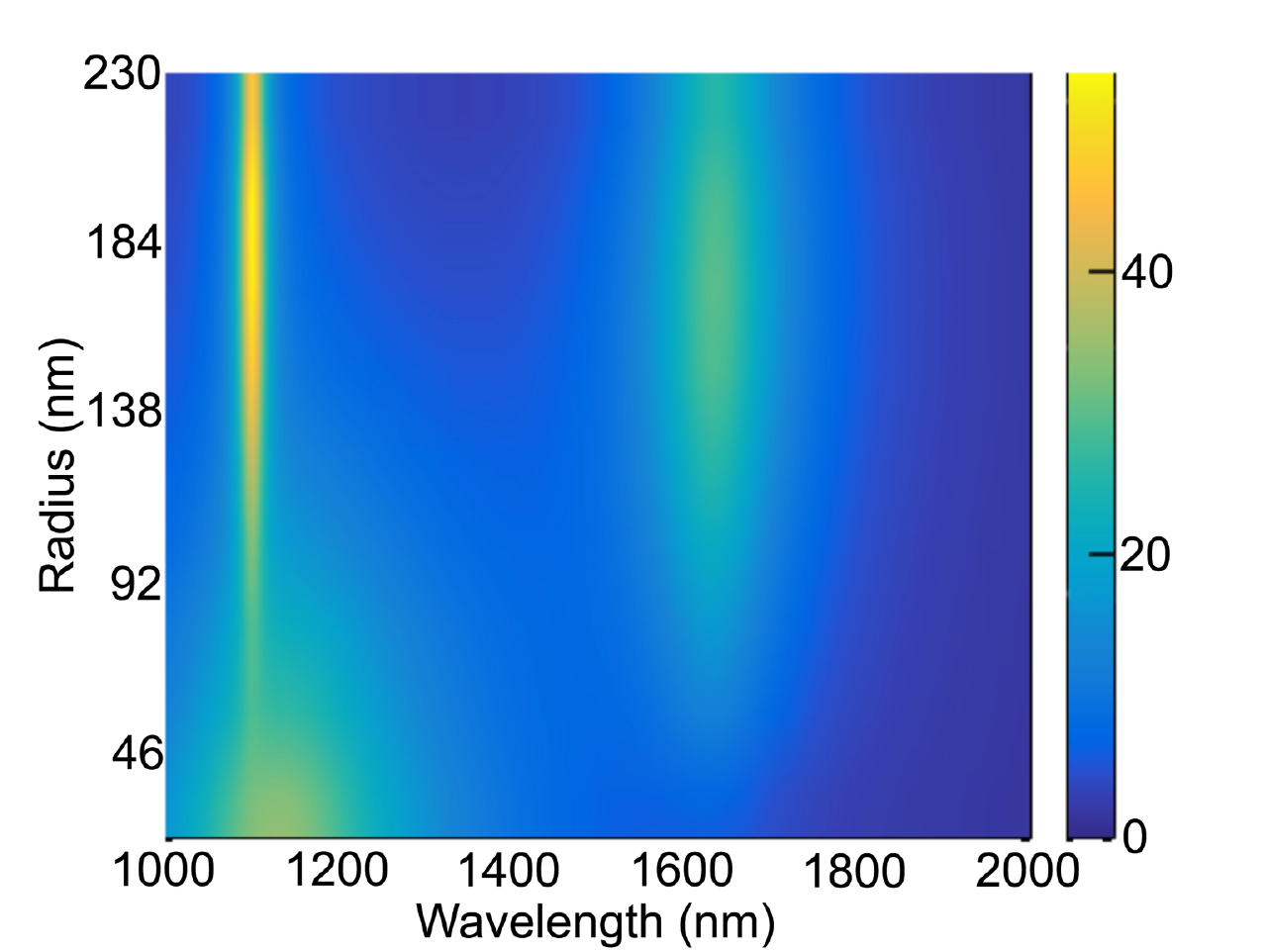}
\caption{Integral of the EM intensity (normalized to the volume) in the core region as a function of core radius and wavelength.}
\label{fig:4}
\end{figure}

Finally, to further verify that the local field intensity in the interior of the sphere actually governs, if not fully, the overall polarization conversion, we present in Fig. \ref{fig:4}  the integral of the EM intensity  in the core region (normalized to its volume) as a function of core radius and wavelength, showing a remarkable agreement with the actual polarization conversion.

In summary we have deeply analyzed the polarization conversion capabilities of high-index, non-absorbing, core-shell dielectric nanoantennas. We have demonstrated that, in analogy with their metallic, plasmonic, counterparts, the polarization conversion is controlled by the internal resonances of the nanosphere. However, in strong contrast with plasmon nanoparticles,  the magneto-optical response is fully governed by the magnetic (dipolar and quadrupolar) resonances with little effect of the electric ones.
We have also pointed out that this behavior arises from the particular spatial field profile within the nanosphere that is, in turn, linked to the nature of each resonance.

\section*{Methods}
Fields and polarisation conversion results were computed using an extended  discrete dipole approximation (DDA) method \cite{draine1994discrete,TsangKongDingEtAl2000,Yurkin2007} for magneto-optical scattering calculations \cite{SmithStokes2006,OsaAlbellaSaizEtAl2010}. We consider a non-homogeneous finite target characterised by a  dielectric permittivity tensor $\ep(\r)$ embedded
in an otherwise homogeneous media with $\epsilon_h =n_{h}^2$ (real).  In absence of free currents, the total electric field is given by the solution of the integral equation
\be
 \E(\r) = \E_0(\r) + k_0^2 \int_V \G(\r,\r')
\left[\bm{\epsilon}(\r')-\epsilon_h {\bf I} \right] \E(\r') d^3\r' 
\label{EqInt}
 \ee
 where ${\bf I}$ is the unit tensor and $\E_0(\r)$ is the solution of the Maxwell equations in absence of the target.
  We define $\G(\r,\r_0)$ as the Green tensor    connecting (through the homogeneous media) an electric-dipole source $\p$ at a position $\r_0$ to the electric field at a position $\r$ 
 by the relation \cite{novotny2012principles} 
 \be \E(\r)_{\text{dipole}} &=& \frac{k^2}{\epsilon_0 \epsilon_h} \G(\r,\r_0)  \p \\
 \G(\r,\r_0)  \p &=& \left[ \p + \frac{[\p.\bm{\nabla}] \bm{\nabla}}{k^2} \right] \frac{e^{ik|\r-\r_0|}}{4\pi|\r-\r_0|} \ee
where $k= \sqrt{\epsilon_h} \ \omega/c $,  $k_0=  \ \omega/c $. 
Following 
 Lakhtakia's \cite{Lakhtakia1992}   theoretical discussion, the DDA is  equivalent to a discretised version of the integral formulation [Eq. \eqref{EqInt}] of the Maxwell equations. 
 The volume of the object, $V$ is considered as the union of non-overlapping, simply connected subregions of volume $V_n$ ($n=1, \dots, N$) with $V=\sum_n V_n$.
     Each subregion $V_n$ is homogeneous and so small that the electric  field can be considered as approximately constant.
 Assuming that $\r_n$ represents a point centred in volume $V_n$ (inside the object), Eq. \eqref{EqInt} can be approximated  as
  \be
 \E(\r) = \E_0(\r) + \frac{k^2}{\epsilon_0 \epsilon_h} \sum_{n=1}^N  \overline{\G}(\r,\r_n)
 \p_n  \quad, \quad 
  \p_n \equiv  \epsilon_0 (\ep(\r_n)-\epsilon_h {\bf I}) V_n \E(\r_n).
    \label{Eext} 
 \ee
 where, $\overline{\G}(\r,\r_n)$ is the Green tensor averaged over $V_n$,
 \be
 k^2\overline{\G}(\r,\r_i) \equiv \frac{k^2}{V_n}  \int_{V_n} \G(\r,\r')
  d^3\r' \approx 
  \begin{cases}
  k^2\G(\r,\r_n) & \text{if} \quad  \r \notin V_n \\
  -{\bf L}_n / V_n + i k^2 \text{Im}\left\{\G(\r_n,\r_n) \right\} =  -{\bf L}_n / V_n +  i k^3/(6\pi) {\bf I}  & \text{if} \quad \r=\r_n
  \end{cases}
  \label{Gapp}
 \ee
 where ${\bf L}_n$ is the electrostatic depolarisation dyadic \cite{Lakhtakia1992,Yaghjian1980}   that depends on the shape of the volume element $V_n$. For a Rectangular parallelepiped of volume $V_n=L_{nx}L_{ny}L_{nz}$, \cite{Yaghjian1980}
 \be
 [{\bf L}_n]_{ij} &=& \delta_{ij} \frac{2}{\pi} \arctan\left\{  \frac{1}{ L^2_{ni} }\frac{V_n}{ \sqrt{L_{nx}^2 + L_{ny}^2 +L_{nz}^2} }\right\} \quad, \quad \sum_{i=1}^3 L_{ii} =1 
 \ee
From Eqs. \eqref{Eext} and \eqref{Gapp} it is easy to find the self-consistent coupled equations for the  internal field, $\E(\r_n)$,
\be
 \frac{1}{\epsilon_h} \left[{\bf I}\epsilon_h + \left({\bf L}_n  -i V_n \frac{k^3}{6\pi} \right)[\ep(\r_n)-\epsilon_h {\bf I}] \right] \E(\r_n) = \E_0(\r_n) +  k^2 \sum_{m \ne n}^N  \G(\r_n,\r_m)
 \frac{\ep(\r_m)-\epsilon_h {\bf I}}{ \epsilon_h} V_m \E(\r_m). \label{EqIntMO2}
 \ee
We can identify the left hand side of equation  \eqref{EqIntMO2} as the 
field, $ \E_{\text{exc}}(\r_n) $, exciting the (dipolar) $n$-volume element.  If we now define the polarizability tensor, $\bm{\alpha}_n$, as
 \be
 \bm{\alpha}_n &\equiv&  \left\{\bm{\alpha}_{n0}^{-1} -i  \frac{k^3}{6\pi}  \right\}^{-1} \\
\bm{\alpha}_{n0} &=& (\ep(\r_n)-\epsilon_h {\bf I})\left[(\ep(\r_n)-\epsilon_h {\bf I}) + {\bf L}_n^{-1} \epsilon_h\right]^{-1} \ {\bf L}_n^{-1} V_n 
\label{alfa0}
\ee
[$\bm{\alpha}_{n0}$ is the quasistatic polarizability tensor], Eq.  \eqref{EqIntMO2}
 can be rewritten as a set of couple dipole equations, for the exciting fields at each element   \be
 \E_{\text{exc}}(\r_n) &=& 
 \E_0(\r_n) +  k^2 \sum_{m \ne n}^N  \G(\r_n,\r_m)  \bm{\alpha}_{m} \  \E_{\text{exc}} (\r_m). \label{EqIntMO3}
 \ee
 Notice that in our approach, the so-called radiative corrections \cite{Sipe1974,Belov2003,albaladejo2010radiative} [related to the imaginary part of the Green Tensor] arise in a natural way and, as a consequence,  the DDA results are found to be fully consistent with the Optical Theorem as discussed below.  For cubic volume elements, like for spheres, the depolarization tensor is diagonal ${\bf L}_n = {\bf I}/3$ and our approach is equivalent to previous extended DDA \cite{OsaAlbellaSaizEtAl2010}.
 
  The numerical solution of the set of $3N$ coupled equations  \eqref{EqIntMO3} give the set of ``exciting'' fields $\left\{ \E_{\text{exc}}(\r_n); n=1,\dots,N\right\}$  from which we get 
  \be
  \p_n &=& \epsilon_0 \epsilon_h \bm{\alpha}_{n} \  \E_{\text{exc}} (\r_n) \label{pnDDA}\\
  \E(\r_n) &=& \frac{1}{\epsilon_0 V_n} (\ep(\r_n)-\epsilon_h {\bf I})^{-1} \p_n.
 \ee
 and,  
 assuming plane wave illumination, $\E_0(\r)= \E_0 e^{i\k.\r}$, 
 the scattering, $\sigma_{\text{scatt}}$, absorption, $\sigma_{\text{abs}}$, and total extinction,  $\sigma_{\text{ext}}$, cross sections can be shown to be given by
 \be
\sigma_{\text{ext}} &=& \frac{k}{\epsilon_0 \epsilon_h |\E_0|^2} \sum_{n=1}^N \text{Im}\left\{ \E_0^*(\r_n).\p_n \right\} \\
\sigma_{\text{scatt}} &=& \frac{k^3}{(\epsilon_0 \epsilon_h)^2 |\E_0|^2} \sum_{n,m=1}^N \p_n^* . \text{Im} \left\{ \G(\r_n,\r_m) \right\} \p_m 
\\
\sigma_{\text{abs}} &=&  \sigma_{\text{ext}}  - \sigma_{\text{scatt}}  \\ &=& \frac{k}{(\epsilon_0 \epsilon_h)^2 |\E_0|^2} \sum_{n=1}^N \Big\{ 
\text{Im}\left\{ \E_0^*(\r_n).\p_n \right\} - k^2
\sum_{m=1}^N \p_n^* .\text{Im} \left\{ \G(\r_n,\r_m) \right\} \p_m 
\Big\}
\\ &=& \frac{k}{\epsilon_0 \epsilon_h |\E_0|^2} \sum_{n=1}^N \text{Im}\left\{
 \p_n^* .\left[ {\bm{\alpha}_{n0}^\dagger}^{-1} \right] \p_n \label{abs}
  \right\}
 \ee
 For a lossless material, the dielectric tensor must be Hermitian $\ep(\r_n)=\ep^\dagger(\r_n) $ and so it is  the inverse of the quasistatic polarizability tensor [${\bm{\alpha}_{n0}^\dagger}^{-1}$]  [Eq. \eqref{alfa0}], which, from Eq. \eqref{abs}, leads do   $\sigma_{\text{abs}} = 0$.


\section*{Acknowledgements}

This research was supported by the Spanish Ministry of Economy and Competitiveness through grants FIS2012-36113-C03, FIS2015-69295-C3-3-P, and
MAT2014-58860-P, and  by the Comunidad de Madrid (Contract No. S2013/MIT-2740). L.S.F.-P. acknowledges funding from the Swiss National Science Foundation through the National Centre of Competence in Research Bio-Inspired Materials.
\section*{Author contributions statement}

All authors contributed equally to this work, participated in the scientific discussions of the results and in the writing of the manuscript. 
\section*{Competing financial interests}
The authors declare no competing financial interests.

\end{document}